\documentstyle[11pt,epsfig]{article}
\title{\bf Anisotropic brane gravity with a confining potential}
\author{M. Heydari-Fard $^{1}$\thanks{email:
m-heydarifard@sbu.ac.ir} and H. R. Sepangi\thanks{email:
hr-sepangi@sbu.ac.ir}
\\ {\small Department of Physics, Shahid Beheshti University, Evin, Tehran 19839, Iran}}
\begin{document}
\maketitle 
\begin{abstract}
We consider an anisotropic brane world with Bianchi type I and V
geometries where the mechanism of confining the matter on the
brane is through the use of a confining potential. The resulting
equations on the anisotropic brane are modified by an extra term
that may be interpreted as the x-matter, providing a possible
phenomenological explanation for the accelerated expansion of the
universe. We obtain the general solution of the field equations in
an exact parametric form for both Bianchi type I and V
space-times. In the special case of a Bianchi type I the solutions
of the field equations are obtained in an exact analytic form.
Finally, we study the behavior of the observationally important
parameters.
\vspace{5mm}\\
\end{abstract}
\section{Introduction}
The type Ia supernovae (SNe Ia) \cite{1} observations provide the
first evidence for the accelerating expansion of the present
universe. These results, when combined with the observations on
the anisotropy spectrum of cosmic microwave background (CMB)
\cite{2} and the results on the power spectrum of large scale
structure (LSS) \cite{3}, strongly suggest that the universe is
spatially flat and dominated by a component, though arguably
exotic, with large negative pressure, referred to as dark energy
\cite{4}. The nature of such dark energy constitutes an open and
tantalizing question connecting cosmology and particle physics.
Different mechanisms have been suggested over the past few years
to accommodate dark energy. The simplest form of dark energy is
the vacuum energy (the cosmological constant). A tiny positive
cosmological constant which can naturally explain the current
acceleration would encounter many theoretical problems such as the
fine-tuning problem and the coincidence problem. The former can be
stated as the existence of an enormous gap between the vacuum
expectation value, in other words the cosmological constant, in
particle physics and that observed over cosmic scales. The absence
of a fundamental mechanism which sets the cosmological constant to
zero or to a very small value is also known as the cosmological
constant problem. The latter however relates to the question of
the near equality of energy densities of the dark energy and dark
matter today.

Another possible form of dark energy is a dynamical, time
dependent and spatially inhomogeneous component, called the
quintessence \cite{5}. An example of quintessence is the energy
associated with a scalar field $\phi$ slowly evolving down its
potential $V(\phi)$ \cite{51,52}. Slow evolution is needed to
obtain a negative pressure,
$p_{\phi}=\frac{1}{2}\dot{\phi}^2+V(\phi)$, so that the kinetic
energy density is less than the potential energy density. Yet
another phenomenological explanation based on current
observational data is given by the x-matter (xCDM) model which is
associated with an exotic fluid characterized by an equation of
state $p_{x}=w_{x}\rho_{x}$ ($w_{x}<-\frac{1}{3}$ is the necessary
condition to make a universe accelerate), where the parameter
$w_{x}$ can be a constant or more generally a function of time
\cite{6}.

Over the past few years, we have been witnessing a phenomenal
interest in the possibility that our observable four-dimensional
($4D$) universe may be viewed as a brane hypersurface embedded in
a higher dimensional bulk space. Physical matter fields are
confined to this hypersurface, while gravity can propagate in the
higher dimensional space-time as well as on the brane. The most
popular model in the context of brane world theory is that
proposed by Randall and Sundrum (RS). In the so-called RSI model
\cite{11}, the authors proposed a mechanism to solve the hierarchy
problem with two branes, while in the RSII model \cite{12}, they
considered a single brane with a positive tension, where $4D$
Newtonian gravity is recovered at low energies even if the extra
dimension is not compact. This mechanism provides us with an
alternative to compactification of extra dimensions. The
cosmological evolution of such a brane universe has been
extensively investigated and effects such as a quadratic density
term in the Friedmann equations have been found
\cite{13}-\cite{15}. This term arises from the imposition of the
Israel junction conditions which is a relationship between the
extrinsic curvature and energy-momentum tensor of the brane and
results from the singular behavior in the energy-momentum tensor.
There has been concerns expressed over applying such junction
conditions in that they may not be unique. Indeed, other forms of
junction conditions exist, so that different conditions may lead
to different physical results \cite{16}. Furthermore, these
conditions cannot be used when more than one non-compact extra
dimension is involved. To avoid such concerns, an interesting
higher-dimensional model was introduced in \cite{17} where
particles are trapped on a 4-dimensional hypersurface by the
action of a confining potential ${\cal V}$. In \cite{18}, the
dynamics of test particles confined to a brane by the action of
such potential at the classical and quantum levels were studied
and the effects of small perturbations along the extra dimensions
investigated. Within the classical limits, test particles remain
stable under small perturbations and the effects of the extra
dimensions are not felt by them, making them undetectable in this
way. The quantum fluctuations of the brane cause the mass of a
test particle to become quantized and, interestingly, the
Yang-Mills fields appear as quantum effects. Also, in \cite{19}, a
braneworld model was studied in which the matter is confined to
the brane through the action of such a potential, rendering the
use of any junction condition unnecessary and predicting a
geometrical explanation for the accelerated expansion of the
universe.

In brane theories the covariant Einstein equations are derived by
projecting the bulk equations onto the brane. This was first done
by Shiromizu, Maeda and Sasaki (SMS) \cite{20} where the
Gauss-Codazzi equations together with Israel junction conditions
were used to obtain the Einstein field equations on the 3-brane.
In a series of recent papers a number of authors \cite{28,33} have
presented detailed descriptions of the dynamic of homogeneous and
anisotropic brane worlds in the SMS model. The study of
anisotropic homogeneous brane world cosmological models has shown
an important difference between these models and standard $4D$
general relativity, namely, that brane universes are born in an
isotropic state. For the anisotropic Bianchi type I and V
geometries, with a conformally flat bulk (vanishing Weyl tensor),
this type of behavior has been found by exactly solving the field
equations \cite{38}.

In this paper, we follow \cite{19} and consider an $m$-dimensional
bulk space without imposing the $Z_2$ symmetry. As mentioned
above, to localize the matter on the brane, a confining potential
is used rather than a delta-function in the energy-momentum
tensor. The resulting equations on the anisotropic brane are
modified by an extra term that may be interpreted as the x-matter,
providing a possible phenomenological explanation for the
accelerated expansion of the universe. The behavior of the
observationally important physical quantities such as anisotropy
and deceleration parameter is studied in this scenario. We should
emphasize here that there is a difference between the model
presented in this work and models introduced in \cite{23,24} in
that in the latter no mechanism is introduced to account for the
confinement of matter to the brane.

\section{The model }
In this section we present a brief review of the model proposed in
\cite{18}.  Consider the background manifold $ \overline{V}_{4} $
isometrically embedded in a pseudo-Riemannian manifold $ V_{m}$ by
the map ${ \cal Y}: \overline{V}_{4}\rightarrow  V_{m} $ such that
\begin{eqnarray}\label{a}
{\cal G} _{AB} {\cal Y}^{A}_{,\mu } {\cal Y}^{B}_{,\nu}=
\bar{g}_{\mu \nu}  , \hspace{.5 cm} {\cal G}_{AB}{\cal
Y}^{A}_{,\mu}{\cal N}^{B}_{a} = 0  ,\hspace{.5 cm}  {\cal
G}_{AB}{\cal N}^{A}_{a}{\cal N}^{B}_{b} = g_{ab}= \pm 1,
\end{eqnarray}
where $ {\cal G}_{AB} $  $ ( \bar{g}_{\mu\nu} ) $ is the metric of
the bulk (brane) space  $  V_{m}  (\overline{V}_{4}) $ in
arbitrary coordinates, $ \{ {\cal Y}^{A} \} $   $  (\{ x^{\mu} \})
$  is the  basis of the bulk (brane) and  ${\cal N}^{A}_{a}$ are
$(m-4)$ normal unit vectors, orthogonal to the brane. Perturbation
of $\bar{V}_{4}$ in a sufficiently small neighborhood of the brane
along an arbitrary transverse direction $\xi$ is given by
\begin{eqnarray}\label{a1}
{\cal Z}^{A}(x^{\mu},\xi^{a}) = {\cal Y}^{A} + ({\cal
L}_{\xi}{\cal Y})^{A}, \label{eq2}
\end{eqnarray}
where $\cal L$ represents the Lie derivative and $\xi^{a}$ $(a =
1,2,...,m-4)$ is a small parameter along ${\cal N}^{A}_{a}$ that
parameterizes the extra noncompact dimensions. By choosing $\xi$
orthogonal to the brane, we ensure gauge independency \cite{18}
and have perturbations of the embedding along a single orthogonal
extra direction $\bar{{\cal N}}_{a}$ giving local coordinates of
the perturbed brane as
\begin{eqnarray}\label{a2}
{\cal Z}^{A}_{,\mu}(x^{\nu},\xi^{a}) = {\cal Y}^{A}_{,\mu} +
\xi^{a}\bar{{\cal N}}^{A}_{a,\mu}(x^{\nu}).
\end{eqnarray}
In a similar manner, one can find that since the vectors
$\bar{{\cal N}}^{A}$ depend only on the local coordinates
$x^{\mu}$, they do not propagate along the extra dimensions. The
above  assumptions lead to the embedding equations of the
perturbed geometry
\begin{eqnarray}\label{a4}
{\cal G}_{\mu \nu }={\cal G}_{AB}{\cal Z}_{\,\,\ ,\mu }^{A}{\cal
Z}_{\,\,\ ,\nu }^{B},\hspace{0.5cm}{\cal G}_{\mu a}={\cal
G}_{AB}{\cal Z}_{\,\,\ ,\mu
}^{A}{\cal N}_{\,\,\ a}^{B},\hspace{0.5cm}{\cal G}_{AB}{\cal N}_{\,\,\ a}^{A}%
{\cal N}_{\,\,\ b}^{B}={g}_{ab}.
\end{eqnarray}
If we set ${\cal N}_{\,\,\ a}^{A}=\delta _{a}^{A}$, the metric of
the bulk space can be written in the following matrix form
\begin{eqnarray}
{\cal G}_{AB}=\left( \!\!\!%
\begin{array}{cc}
g_{\mu \nu }+A_{\mu c}A_{\,\,\nu }^{c} & A_{\mu a} \\
A_{\nu b} & g_{ab}%
\end{array}%
\!\!\!\right) ,  \label{F}
\end{eqnarray}%
where
\begin{eqnarray}
g_{\mu \nu }=\bar{g}_{\mu \nu }-2\xi ^{a}\bar{K}_{\mu \nu a}+\xi ^{a}\xi ^{b}%
\bar{g}^{\alpha \beta }\bar{K}_{\mu \alpha a}\bar{K}_{\nu \beta
b}, \label{G}
\end{eqnarray}%
is the metric of the perturbed brane, so that
\begin{eqnarray}
\bar{K}_{\mu \nu a}=-{\cal G}_{AB}{\cal Y}_{\,\,\,,\mu }^{A}{\cal
N}_{\,\,\ a;\nu }^{B},  \label{H}
\end{eqnarray}%
represents the extrinsic curvature of the original brane (second
fundamental form). We use the notation $A_{\mu c}=\xi ^{d}A_{\mu
cd}$, where
\begin{equation}
A_{\mu cd}={\cal G}_{AB}{\cal N}_{\,\,\ d;\mu }^{A}{\cal N}_{\,\,\ c}^{B}=%
\bar{A}_{\mu cd},  \label{I}
\end{equation}%
represents the twisting vector fields
(the normal fundamental form). Any fixed $%
\xi ^{a}$ signifies a new perturbed geometry, enabling us to
define an extrinsic curvature similar to the original one by
\begin{eqnarray}
\widetilde{K}_{\mu \nu a}=-{\cal G}_{AB}{\cal Z}_{\,\,\ ,\mu }^{A}{\cal N}%
_{\,\,\ a;\nu }^{B}=\bar{K}_{\mu \nu a}-\xi ^{b}\left( \bar{K}_{\mu \gamma a}%
\bar{K}_{\,\,\ \nu b}^{\gamma }+A_{\mu ca}A_{\,\,\ b\nu
}^{c}\right) . \label{J}
\end{eqnarray}%
Note that definitions (\ref{F}) and (\ref{J}) require
\begin{eqnarray}
\widetilde{K}_{\mu \nu a}=-\frac{1}{2}\frac{\partial {\cal G}_{\mu \nu }}{%
\partial \xi ^{a}}.  \label{M}
\end{eqnarray}%
In geometric language, the presence of gauge fields $A_{\mu a}$
tilts the embedded family of sub-manifolds with respect to the
normal vector ${\cal N} ^{A}$. According to our construction, the
original brane is orthogonal to the normal vector ${\cal N}^{A}.$
However,  equation (\ref{a4})  shows that this is not true for the
deformed geometry. Let us change the embedding coordinates and set
\begin{eqnarray}
{\cal X}_{,\mu }^{A}={\cal Z}_{,\mu }^{A}-g^{ab}{\cal
N}_{a}^{A}A_{b\mu }. \label{mama40}
\end{eqnarray}%
The coordinates ${\cal X}^{A}$ describe a new family of embedded
manifolds whose members are always orthogonal to ${\cal N}^{A}$.
In this coordinates the embedding equations of the perturbed brane
is similar to the original one, described by equation (\ref{a}),
so that ${\cal Y}^{A}$ is replaced by ${\cal X}^{A}$. This new
embedding of the local coordinates is suitable for obtaining
induced Einstein field equations on the brane. The extrinsic
curvature of a perturbed brane then becomes
\begin{eqnarray}
K_{\mu \nu a}=-{\cal G}_{AB}{\cal X}_{,\mu }^{A}{\cal N}_{a;\nu }^{B}=\bar{K}%
_{\mu \nu a}-\xi ^{b}\bar{K}_{\mu \gamma a}\bar{K}_{\,\,\nu b}^{\gamma }=-%
\frac{1}{2}\frac{\partial g_{\mu \nu }}{\partial \xi ^{a}},
\label{mama42}
\end{eqnarray}%
which is the generalized York's relation and shows how the
extrinsic curvature propagates as a result of the propagation of
the metric in the direction of extra dimensions. The components of
the Riemann tensor of the bulk written in the embedding vielbein
$\{{\cal X}^{A}_{, \alpha}, {\cal N}^A_a \}$, lead to the
Gauss-Codazzi equations \cite{27}
\begin{eqnarray}\label{a5}
R_{\alpha \beta \gamma \delta}=2g^{ab}K_{\alpha[ \gamma
a}K_{\delta] \beta b}+{\cal R}_{ABCD}{\cal X} ^{A}_{,\alpha}{\cal
X} ^{B}_{,\beta}{\cal X} ^{C}_{,\gamma} {\cal X}^{D}_{,\delta},
\end{eqnarray}
\begin{eqnarray}\label{ab5}
2K_{\alpha [\gamma c; \delta]}=2g^{ab}A_{[\gamma ac}K_{ \delta]
\alpha b}+{\cal R}_{ABCD}{\cal X} ^{A}_{,\alpha} {\cal N}^{B}_{c}
{\cal X} ^{C}_{,\gamma} {\cal X}^{D}_{,\delta},
\end{eqnarray}
where ${\cal R}_{ABCD}$ and $R_{\alpha\beta\gamma\delta}$ are the
Riemann tensors for the bulk and the perturbed brane respectively.
Contracting the Gauss equation (\ref{a5}) on ${\alpha}$ and
${\gamma}$, we find
\begin{eqnarray}\label{a7}
R_{\mu\nu}=(K_{\mu\alpha c}K_{\nu}^{\,\,\,\,\alpha c}-K_{c} K_{\mu
\nu }^{\,\,\,\ c})+{\cal R}_{AB} {\cal X}^{A}_{,\mu} {\cal
X}^{B}_{,\nu}-g^{ab}{\cal R}_{ABCD}{\cal N}^{A}_{a}{\cal
X}^{B}_{,\mu}{\cal X}^{C}_{,\nu}{\cal N}^{D}_{b},
\end{eqnarray}
and the Einstein tensor of the brane is given by
\begin{eqnarray}\label{a8}
G_{\mu\nu}=G_{AB} {\cal X}^{A}_{,\mu}{\cal
X}^{B}_{,\nu}+Q_{\mu\nu}+g^{ab}{\cal R}_{AB}{\cal N}^{A}_{a}{\cal
N}^{B}_{b} g_{\mu\nu}- g^{ab}{\cal R}_{ABCD}{\cal N}^{A}_{a}{\cal
X}^{B}_{,\mu}{\cal X}^{C}_{,\nu}{\cal N}^{D}_{b},
\end{eqnarray}
where
\begin{eqnarray}\label{a9}
Q_{\mu\nu}=-g^{ab}\left(K^\gamma_{\mu a}K_{\gamma\nu b}-K_a
K_{\mu\nu b}\right)+\frac{1}{2}\left(K_{\alpha\beta
a}K^{\alpha\beta a}-K_a K^a\right)g_{\mu\nu}. \label{eqq7}
\end{eqnarray}
As can be seen from the definition of $Q_{\mu\nu}$,  it is
independently a conserved quantity, that is
$Q^{\mu\nu}_{\,\,\,\,\, ;\nu}=0$ \cite{23}. Using the
decomposition of the Riemann tensor into the Weyl curvature, the
Ricci tensor and the scalar curvature
\begin{eqnarray}\label{a10}
{\cal R}_{ABCD}=C_{ABCD}-\frac{2}{(m-2)}\left({\cal G}_{B[D}{\cal
R}_{C]A}-{\cal G}_{A[D}{\cal
R}_{C]B}\right)-\frac{2}{(m-1)(m-2)}{\cal R}({\cal G}_{A[D}{\cal
G}_{C]B}),
\end{eqnarray}
we obtain the $4D$ Einstein equations as
\begin{eqnarray}\label{a11}
G_{\mu\nu}&=&G_{AB} {\cal X}^{A}_{,\mu}{\cal
X}^{B}_{,\nu}+Q_{\mu\nu}-{\cal
E}_{\mu\nu}+\frac{m-3}{(m-2)}g^{ab}{\cal R}_{AB}{\cal
N}^{A}_{a}{\cal
N}^{B}_{b}g_{\mu\nu}\nonumber\\
&-&\frac{m-4}{(m-2)}{\cal R}_{AB}{\cal X}^{A}_{,\mu}{\cal
X}^{B}_{,\nu}+\frac{m-4}{(m-1)(m-2)}{\cal
R}g_{\mu\nu},\label{eqq12}
\end{eqnarray}
where
\begin{eqnarray}\label{a12}
{\cal E}_{\mu\nu}=g^{ab} C_{ABCD}{\cal N}^{A}_{a}{\cal
X}^{B}_{,\mu}{\cal N}^D_b{\cal X}^C_{,\nu},
\end{eqnarray}
is the electric part of the Weyl tensor $C_{ABCD}$. Now, let us
write the Einstein equation in the bulk space as
\begin{eqnarray}\label{a13}
G^{(b)}_{AB}+\Lambda^{(b)} {\cal G}_{AB}=\alpha^{*}
S_{AB},\label{eqq14}
\end{eqnarray}
where
\begin{eqnarray}\label{a14}
S_{AB}=T_{AB}+ \frac{1}{2} {\cal V} {\cal G}_{AB},\label{eqq15}
\end{eqnarray}
here $\alpha^{*}=\frac{1}{M_{*}^{m-2}}$ ($M_{*}$ is the
fundamental scale of energy in the bulk space), $\Lambda^{(b)}$ is
the cosmological constant of the bulk and $T_{AB}$ is the
energy-momentum tensor of the matter confined to the brane through
the action of the confining potential $\cal{V}$. We require
$\cal{V}$  to satisfy three general conditions: firstly, it has a
deep minimum on the non-perturbed brane, secondly, depends only on
extra coordinates and thirdly, the gauge group representing the
subgroup of the isometry group of the bulk space is preserved by
it \cite{18}. The vielbein components of the energy-momentum
tensor are given by
\begin{eqnarray}\label{aa14}
S_{\mu\nu}=S_{AB}{\cal X}^{A}_{,\mu}{\cal X}^{B}_{,\nu},\hspace{.5
cm} S_{\mu a}=S_{AB}{\cal X}^{A}_{,\mu}{\cal N}^{B}_{a},\hspace{.5
cm} S_{ab}=S_{AB}{\cal N}^{A}_{a}{\cal N}^{B}_{b}.
\end{eqnarray}
Use of equation (\ref{eqq15}) then gives
\begin{eqnarray}\label{a15}
{\cal R}_{AB}=-\frac{\alpha^{*}}{(m-2)}{\cal G}_{AB}
S+\frac{2}{(m-2)}\Lambda^{(b)} {\cal G}_{AB}+\alpha^{*} S_{AB},
\end{eqnarray}
and
\begin{eqnarray}\label{a16}
{\cal R}=-\frac{2}{m-2}(\alpha^{*} S-m\Lambda^{(b)}).
\end{eqnarray}
Substituting ${\cal R}_{AB}$ and ${\cal R}$ from the above into
equation (\ref{a11}) and using equation (\ref{aa14}), we obtain
\begin{eqnarray}\label{a17}
G_{\mu\nu}&=& Q_{ \mu\nu} - {\cal
E}_{\mu\nu}+\frac{(m-3)}{(m-2)}\alpha^{*}g^{ab}S_{ab}g_{\mu\nu}
+\frac{2\alpha^{*}}{(m-2)}S_{\mu\nu} -
\frac{(m-4)(m-3)}{(m-1)(m-2)}\alpha^{*}Sg_{\mu\nu}\nonumber\\
&+&\frac{(m-7)}{(m-1)}\Lambda^{(b)}g_{\mu\nu}.\label{new1}
\end{eqnarray}
On the other hand, again from equation (\ref{eqq14}), the trace of
the Codazzi equation (\ref{ab5}) gives the ``gravi-vector
equation''
\begin{equation}
K^\delta_{a\gamma;\delta} - K_{a,\gamma} - A_{ba\gamma}K^b +
A_{ba\delta}K^{b\delta} + B_{a\gamma} =
\frac{3(m-4)}{m-2}\alpha^{*}S_{a\gamma},\label{new2}
\end{equation}
where
\begin{equation}
B_{a\gamma} = g^{mn}C_{ABCD}{\cal N}^A_m{\cal N}^B_a{\cal
X}^C_{,\gamma}{\cal N}^D_n.
\end{equation}
Finally, the ``gravi-scalar equation'' is obtained from the
contraction of (\ref{a7}), (\ref{a11}) and using equation
(\ref{a13})
\begin{equation}
\alpha^{*}\left[\frac{m-5}{m-1}S - g^{mn}S_{mn}\right]g_{ab} =
\frac{m-2}{6}\left(Q + R +W\right)g_{ab} -
\frac{4}{m-1}\Lambda^{(b)}g_{ab},\label{new3}
\end{equation}
where
\begin{equation}
W = g^{ab}g^{mn}C_{ABCD}{\cal N}^A_m{\cal N}^B_b{\cal N}^C_n{\cal
N}^D_a.\label{New3}
\end{equation}
Equations (\ref{new1})-(\ref{New3}) represent the projections of
the Einstein field equations on the brane-brane, bulk-brane, and
bulk-bulk directions.

As was mentioned in the introduction, localization of matter on
the brane is realized in this model by the action of a confining
potential. Since the potential $\cal V$ is assumed to have a
minimum on the brane, which can be taken as zero, localization of
matter may simply be realized by taking equation (\ref{eqq15}) and
consider its components on the brane, in which case we may write
\begin{eqnarray}
\alpha\tau_{\mu\nu} = \frac{2\alpha^{*}}{(m-2)}T_{\mu\nu},
\hspace{.5 cm}T_{\mu a}=0, \hspace{.5 cm}T_{ab}=0,\label{new4}
\end{eqnarray}
where $\alpha$ is the scale of energy on the brane. Now, the
induced Einstein field equations on the original brane can be
written as
\begin{eqnarray}
G_{\mu\nu} = \alpha \tau_{\mu\nu}
-\frac{(m-4)(m-3)}{2(m-1)}\alpha\tau g_{\mu\nu} - \Lambda
g_{\mu\nu} + Q_{\mu\nu} - {\cal E}_{\mu\nu},\label{a8}
\end{eqnarray}
where  $\Lambda= -\frac{(m-7)}{(m-1)} \Lambda^{(b)}$ and
$Q_{\mu\nu}$ is a conserved quantity which according to \cite{23}
may be considered as an energy-momentum tensor of a dark energy
fluid representing the x-matter, the more common phrase being
`x-Cold-Dark Matter' (xCDM). This matter has the most general form
of the equation of state which is characterized by the following
conditions \cite{25}:  violation of the strong energy condition at
the present epoch for $\omega_x<-1/3$ where $p_x=\omega_x\rho_x$,
local stability {\it i.e.} $c^2_s=\delta p_x/\delta\rho_x\ge 0$
and preservation of causality {\it i.e.} $c_s\le 1$. Ultimately,
we have three different types of `matter' on the right hand side
of equation (\ref{a8}), namely, ordinary confined conserved matter
represented by $\tau_{\mu\nu}$, the matter represented by
$Q_{\mu\nu}$ which will be discussed later and finally, the Weyl
matter represented by ${\cal E}_{\mu\nu}$.
\section{Field equations on the anisotropic brane}
In the following we will investigate the influence of the
extrinsic curvature terms on the anisotropic universe described by
Bianchi type I and V geometries. We restrict our analysis to a
constant curvature bulk, so that ${\cal E}_{\mu\nu}=0$. The
constant curvature bulk is characterized by the Riemann tensor
\begin{eqnarray}
{\cal R}_{ABCD}=k_{*}({\cal G}_{AC}{\cal G}_{BD}-{\cal
G}_{AD}{\cal G}_{BC}),\label{1}
\end{eqnarray}
where ${\cal G}_{AB}$ denotes the bulk metric components in
arbitrary coordinates and $k_{*}$ is either zero for the flat
bulk, or proportional to a positive or negative bulk cosmological
constant respectively, corresponding to two possible signatures
$(4,1)$ for the $dS_{5}$ bulk and $(3,2)$ for the $AdS_{5}$ bulk.
We take, in the embedding equations, $g^{55}=\varepsilon=\pm1$.
With this assumption the Gauss-Codazzi equations reduce to
\begin{eqnarray}
R_{\alpha\beta\gamma\delta} = \frac{1}{\varepsilon}
(K_{\alpha\gamma}K_{\beta\delta}-K_{\alpha\delta}K_{\beta\gamma})
+ k_{*}
(g_{\alpha\gamma}g_{\beta\delta}-g_{\alpha\delta}g_{\beta\gamma}),\label{2}
\end{eqnarray}
\begin{eqnarray}
K_{\alpha[\beta;\gamma]} = 0.\label{3}
\end{eqnarray}
The effective equations derived in the previous section with
constant curvature bulk can be written as
\begin{eqnarray}
G_{\mu\nu} = \alpha \tau_{\mu\nu}-\lambda g_{\mu\nu} +
Q_{\mu\nu}.\label{4}
\end{eqnarray}
Here, $\lambda$ is the effective cosmological constant in four
dimensions with $Q_{\mu\nu}$ being a completely geometrical
quantity given by
\begin{eqnarray}
Q_{\mu\nu}=\frac{1}{\varepsilon}\left[ \left(KK_{\mu\nu}-
K_{\rho\alpha }K^{\alpha}_{\nu}\right)+\frac{1}{2}
\left(K_{\alpha\beta}K^{\alpha\beta}-K^2\right)g_{\mu\nu}\right],\label{5}
\end{eqnarray}
where $K=g^{\mu\nu}K_{\mu\nu}$. To proceed further, the confined
source on the brane should now be specified. The most common
matter source used in cosmology is that of a perfect fluid which,
in co-moving coordinates, is given by
\begin{eqnarray}
\tau_{\mu\nu}=(\rho+p )u_{\mu}u_{\nu}+p g_{\mu\nu},\hspace{.5
cm}u_{\mu}=-\delta^{0}_{\mu},\hspace{.5
cm}p=(\gamma-1)\rho,\hspace{.5 cm}1\leq\gamma\leq2,\label{6}
\end{eqnarray}
where $\gamma=2$ represents the stiff cosmological fluid
describing the high energy density regime of the early universe.

From a formal point of view the Bianchi type I and V geometries
are described by the line element
\begin{eqnarray}
ds^2 = - dt^2 + a_1^2(t) dx^2 + a_2^2(t)e^{-2\beta x} dy^2 +
a_3^2(t) e^{-2\beta x} dz^2.\label{7}
\end{eqnarray}
The  metric for the Bianchi type I geometry corresponds to the
case $\beta=0$, while for the Bianchi type V case we have
$\beta=1$. Here $a_i(t), i=1,2,3$ are the expansion factors in
different spatial directions. For later convenience, we define the
following variables \cite{38}
\begin{eqnarray}
v=\prod_{i=1}^3 a_i,\hspace{.5 cm} H_i=\frac{\dot
a_i}{a_i},i=1,2,3,\hspace{.5 cm} 3H=\sum_{i=1}^3 H_i,\hspace{.5
cm} \Delta H_i &=& H_i - H,i=1,2,3.\label{8}
\end{eqnarray}
In equation (\ref{8}), $v$ is the volume scale factor, $H_{i},
i=1,2,3$ are the directional Hubble parameters, and $H$ is the
mean Hubble parameter. From equation (\ref{8}) we also obtain
$H=\frac{\dot{v}}{3v}$. The physical quantities of observational
importance in cosmology are the expansion scalar $\Theta$, the
mean anisotropy parameter $A$, and the deceleration parameter $q$,
which are defined according to
\begin{eqnarray}
\Theta=3H,\hspace{.5 cm}3A=\sum_{i=1}^3 \left( \frac{\Delta
H_i}{H} \right)^2,\hspace{.5 cm}q=\frac{d}{d t} H^{-1}-1= -H^{-2}
\left(\dot H + H^2\right).\label{9}
\end{eqnarray}
We note that $A=0$ for an isotropic expansion. Moreover, the sign
of the deceleration parameter indicates how the universe expands.
A positive sign for $q$ corresponds to the standard decelerating
models whereas a negative sign indicates an accelerating expansion
in late times.

Using the York's relation
\begin{eqnarray}
K_{\mu \nu a}=-\frac{1}{2}\frac{\partial
g_{\mu\nu}}{\partial\xi^{a}},\label{10}
\end{eqnarray}
we realize that in a diagonal metric, $K_{\mu\nu a}$ is diagonal.
After separating the spatial components, the Codazzi equations
reduce to (here $\alpha,\beta,\gamma,\sigma=1,2,3$)
\begin{eqnarray}
K^{\alpha}_{\,\,\,\gamma a,\sigma}+K^{\beta}_{\,\,\,\gamma
a}\Gamma^{\alpha}_{\,\,\,\beta\sigma}= K^{\alpha}_{\,\,\,\sigma
a,\gamma}+K^{\beta}_{\,\,\,\sigma
a}\Gamma^{\alpha}_{\,\,\,\beta\gamma},\label{11}
\end{eqnarray}
\begin{eqnarray}
K^{\alpha}_{\,\,\,\gamma
a,0}+\frac{\dot{a_i}}{a_i}K^{\alpha}_{\,\,\,\gamma
a}=\frac{\dot{a_i}}{a_i}K^{0}_{\,\,\,0a},\hspace{.5
cm}i=1,2,3.\label{12}
\end{eqnarray}
The first equation gives $K^{1}_{\,\,\,1 a,\sigma}=0$  for
$\sigma\neq1$, since $K^{1}_{\,\,\,1 a}$ does not depend on the
spatial coordinates. Repeating the same procedure for
$\alpha,\gamma=i, i=2,3,$ we obtain $K^{2}_{\,\,\,2 a,\sigma}=0$
for $\sigma\neq2$ and $K^{3}_{\,\,\,3 a,\sigma}=0$ for
$\sigma\neq3$. Assuming $K^{1}_{\,\,\,1 a}=K^{2}_{\,\,\,2
a}=K^{3}_{\,\,\,3 a}=b_{a}(t)$, where $b_{a}(t)$ are arbitrary
functions of $t$, the second equation gives
\begin{eqnarray}
\dot{b_{a}}+\frac{\dot{a}_i}{a_i}b_{a}=\frac{\dot{a}_i}{a_i}K^{0}_{\,\,\,0
a},\hspace{.5 cm}i=1,2,3.\label{13}
\end{eqnarray}
Summing equations (\ref{13}) we find
\begin{eqnarray}
K_{00
a}=-\left(\frac{3\dot{b}_{a}v}{\dot{v}}+b_{a}\right).\label{14}
\end{eqnarray}
For $\mu,\nu=1,2,3 $ we obtain
\begin{eqnarray}
K_{\mu\nu a}=b_{a}g_{\mu\nu}.\label{15}
\end{eqnarray}
Assuming further that the functions $b_{a}$ are equal and denoting
$b_{a}=b$, $\theta=\frac{\dot{b}}{b}$ and
$\Theta=\frac{\dot{v}}{v}$, we find from equation (\ref{5}) that
\begin{eqnarray}
K_{\alpha\beta}
K^{\alpha\beta}=b^2\left(9\frac{\theta^2}{\Theta^2}+6\frac{\theta}{\Theta}+4\right)
,\hspace{.5
cm}K=b\left(3\frac{\theta}{\Theta}+4\right),\label{166}
\end{eqnarray}
\begin{eqnarray}
Q_{\mu\nu}=-\frac{3
b^2}{\varepsilon}\left(\frac{2\theta}{\Theta}+1\right)g_{\mu\nu},
\hspace{5mm}\mu,\nu=1,2,3,\hspace{.5 cm} Q_{00}=\frac{3
b^2}{\varepsilon}.\label{16}
\end{eqnarray}
Now, using these relations and equation (\ref{4}), the modified
Friedmann equations become
\begin{eqnarray}
3 \dot H + \sum_{i=1}^3 H_i^2=\lambda -
\frac{\alpha}{2}\rho(3\gamma-2)+\frac{3b^2}{\varepsilon}\left
(\frac{3v\dot{b}}{\dot{v}b}+1\right),\label{17}
\end{eqnarray}
\begin{eqnarray}
\frac{1}{v} \frac{d}{d t} (v H_i)=2\beta^2 v^{-2/3}+\lambda
-\frac{\alpha}{2}\rho(\gamma-2)
+\frac{3b^2}{\varepsilon}\left(\frac{v\dot{b}}{\dot{v}b}+1\right),\hspace{.5
cm}i=1,2,3.\label{18}
\end{eqnarray}
For $\beta=0$ we obtain the field equations for Bianchi type I
geometry, while $\beta=1$ gives the Bianchi type V equations on
the brane world. These equations are modified with respect to the
standard equations by the components of the extrinsic curvature.
Such term may be used to offer an explanation for the x-matter. In
the next section we discuss the ramifications of this term on the
cosmology of our model. We also note the implicit effects of the
bulk signature $\varepsilon$ on the expansion of the universe.

For the sake of completeness, let us compare the model presented
in this work to the usual brane worlds models where the Israel
junction condition is used to calculate the extrinsic curvature in
terms of the energy-momentum tensor on the brane and its trace,
that is
\begin{eqnarray}
K_{\mu\nu}=-\frac{1}{2}\alpha^{* 2}(\tau_{\mu\nu}-\frac{1}{3}\tau
g_{\mu\nu}),\label{188}
\end{eqnarray}
where $\alpha^{*}$ is proportional to the gravitational constant
in the bulk. If we did that, we would obtain
$b(t)=-\frac{1}{6}\alpha^{* 2}\rho$, which, upon its substitution
in equation (\ref{17}), gives
\begin{eqnarray}
3 \dot H + \sum_{i=1}^3 H_i^2=\lambda +
\frac{\alpha}{2}\rho(2-3\gamma)+\frac{\alpha^{*
 4}}{12\varepsilon}\rho^2(1-3\gamma),\label{189}
\end{eqnarray}
which is same as equation (16) in \cite{38}. Therefore, the
emergence of a $\rho^2$ term in the Friedmann equations is a
consequence of the discontinuity in the bulk and the brane system.
The existence of this term either does not agree with observations
or requires extra parameters and fine tuning.
\section{Dark energy and role of extrinsic curvature}
As we noted before, $Q_{\mu\nu}$ is an independently conserved
quantity, suggesting that an analogy with the energy momentum of
an uncoupled non-conventional energy source would be in order. To
evaluate the compatibility of such geometrical model with the
present experimental data, we identify $Q_{\mu\nu}$ with x-matter
\cite{25} by defining $Q_{\mu\nu}$ as a perfect fluid and write
\begin{eqnarray}
Q_{\mu\nu}=\frac{1}{\alpha}\left[(\rho_{x}+p_{x})
u_{\mu}u_{\nu}+p_{x} g_{\mu\nu}\right] ,\hspace{.5 cm}
p_{x}=(\gamma_{x}-1)\rho_{x}.\label{19}
\end{eqnarray}
Comparing $Q_{\mu\nu}, \mu,\nu=1,2,3$ and $Q_{00}$ from equation
(\ref{19}) with the components of $Q_{\mu\nu}$ and $Q_{00}$ given
by equation (\ref{16}), we obtain
\begin{eqnarray}\
p_{x}=-\frac{3b^2}{\alpha
\varepsilon}\left(\frac{2\theta}{\Theta}+1\right) ,\hspace{.5
cm}\rho_{x}=\frac{3b^2}{\alpha\varepsilon}.\label{20}
\end{eqnarray}
Equation (\ref{19}) was chosen in accordance with the weak-energy
condition corresponding to the positive energy density and
negative pressure with $\varepsilon=+1$. Use of the above
equations leads to an equation for $b(t)$
\begin{eqnarray}
\frac{\dot{b}}{b}=-\frac{\gamma_{x}}{2}\left(\frac{\dot{v}}{v}\right).\label{21}
\end{eqnarray}
If $\gamma_{x}$ is taken as a constant, the solution for $b(t)$ is
\begin{eqnarray}
b(t)=b_{0}v^{-\gamma_{x}/{2}}.\label{22}
\end{eqnarray}
Using equation (\ref{20}) and this solution, the energy density of
xCDM becomes
\begin{eqnarray}
\rho_{x}=\frac{3b_{0}^{2}}{\alpha}v^{-\gamma_{x}}.\label{23}
\end{eqnarray}

A brief discussion on the energy-momentum conservation on the
brane would be appropriate at this point. The contracted Bianchi
identities in the bulk space, $G^{AB}_{\,\,\,\,\,\,\,\,\,;A}=0$,
using equation (\ref{eqq14}), imply
\begin{eqnarray}
\left(T^{AB}+\frac{1}{2} {\cal{V}} {\cal {G}}^{AB}\right)_{
;A}=0.\label{24}
\end{eqnarray}
Since the potential $\cal V$ has a minimum on the brane, the above
conservation equation reduces to
\begin{eqnarray}
\tau^{\mu\nu}_{\,\,\,\,\,;\mu}=0,\label{25}
\end{eqnarray}
and gives
\begin{eqnarray}
\dot{\rho}+\frac{\dot{v}}{v}\gamma\rho=0.\label{26}
\end{eqnarray}
Thus, the time evolution of the energy density of the matter is
given by
\begin{eqnarray}
\rho=\rho_{0}v^{-\gamma}.\label{27}
\end{eqnarray}
Using the geometrical energy density for $Q_{\mu\nu}$ and the
evolution law of the matter energy density, the field equations
(\ref{17})-(\ref{18}) become
\begin{eqnarray}
3 \dot H + \sum_{i=1}^3 H_i^2=\lambda +
\frac{\alpha}{2}\rho_{0}(2-3\gamma)v^{-\gamma}+\frac{3}{2}b_{0}^2(2-3\gamma_{x})
v^{-\gamma_{x}},\label{28}
\end{eqnarray}
\begin{eqnarray}
\frac{1}{v} \frac{d}{d t} (v H_i)=2\beta^2 v^{-2/3}+\lambda
+\frac{\alpha}{2}\rho_{0}(2-\gamma)v^{-\gamma}
+\frac{3}{2}b_{0}^2(2-\gamma_{x})v^{-\gamma_{x}},\hspace{.5
cm}i=1,2,3.\label{29}
\end{eqnarray}
Summing equations (\ref{29}) we find
\begin{eqnarray}
\frac{1}{v} \frac{d}{d t} (v H)=2\beta^2 v^{-2/3}+\lambda
+\frac{\alpha}{2}\rho_{0}(2-\gamma)v^{-\gamma}
+\frac{3}{2}b_{0}^2(2-\gamma_{x})v^{-\gamma_{x}}.\label{30}
\end{eqnarray}
Now, substituting back equation (\ref{30}) into equations
(\ref{29}) we obtain
\begin{eqnarray}
H_i = H + \frac{h_i}{v}, \qquad i=1,2,3,\label{31}
\end{eqnarray}
with $h_i, \, i=1,2,3$ being constants of integration satisfying
the consistency condition $\sum_{i=1}^3 h_i=0$. The basic equation
describing the dynamics of the anisotropic brane world with a
constant curvature bulk can be written as
\begin{eqnarray}
\ddot v=6\beta^2v^{1/3}+3 \lambda v +\frac{3\alpha}{2}
\rho_{0}(2-\gamma)v^{1-\gamma}
+\frac{9}{2}b_{0}^2(2-\gamma_{x})v^{1-\gamma_{x}}.\label{32}
\end{eqnarray}
Here, we note that for a stiff fluid $(\gamma=2)$, the dynamics of
the matter on the brane is solely determined by the geometrical
matter (xCDM). The general solution of equation (\ref{32}) becomes
\begin{eqnarray}
t - t_0 = \int \left(9\beta^{2}v^{4/3}+3\lambda v^2+3\alpha
\rho_{0}v^{2-\gamma}
+9b_{0}^{2}v^{2-\gamma_{x}}+C\right)^{-1/2}dv,\label{33}
\end{eqnarray}
where $C$ is a constant of integration. The time variation of the
physically important parameters described above in the exact
parametric form, with $v$ taken as a parameter, is given by
\begin{eqnarray}
\Theta=3H=\frac{\left(9\beta^{2}v^{4/3}+3\lambda v^2+3\alpha
\rho_{0}v^{2-\gamma}
+9b_{0}^{2}v^{2-\gamma_{x}}+C\right)^{1/2}}{v},\label{35}
\end{eqnarray}
\begin{eqnarray}
a_i=a_{0i} v^{1/3} \exp \left[ h_i \int v^{-1}
\left(9\beta^{2}v^{4/3}+3\lambda v^2+3\alpha \rho_{0}v^{2-\gamma}
+9b_{0}^{2}v^{2-\gamma_{x}}+C\right)^{-1/2} dv \right], \quad
i=1,2,3,\label{37}
\end{eqnarray}
\begin{eqnarray}
A=3h^2 \left(9\beta^{2}v^{4/3}+3\lambda v^2+3\alpha
\rho_{0}v^{2-\gamma}
+9b_{0}^{2}v^{2-\gamma_{x}}+C\right)^{-1},\label{36}
\end{eqnarray}
\begin{eqnarray}
q=\frac{9\beta^2v^{4/3}+\frac{9\alpha\gamma}{2}\rho_{0}v^{2-\gamma}
+\frac{27\gamma_{x}}{2}b_{0}^2v^{2-\gamma_{x}}+3C}{
\left(9\beta^{2}v^{4/3}+3\lambda v^2+3\alpha \rho_{0}v^{2-\gamma}
+9b_{0}^{2}v^{2-\gamma_{x}}+C\right)}-1,\label{38}
\end{eqnarray}
where $h^2=\sum_{i=1}^3 h_i^2$. In addition, the integration
constants $h_{i}$ and $C$ must satisfy the consistency condition
$h^2=\frac{2}{3}C$. For $\beta=0$ we obtain the general solutions
for Bianchi type I geometry, while $\beta=1$ gives the Bianchi
type V solutions on the anisotropic brane world.

In a matter dominated Bianchi type I universe, $\gamma=1$ with
$\gamma_{x}=0$, equation (\ref{33}) becomes
\begin{eqnarray}
t - t_0 = \int
\frac{dv}{\sqrt{3\alpha\rho_{0}v+9b_{0}^2v^2+C}},\label{39}
\end{eqnarray}
where for later convenience, we take
$C=\frac{3\alpha^2\varrho_{0}^2}{12b_{0}^2}$. The time dependence
of the volume scale factor of the Bianchi type I universe is given
by
\begin{eqnarray}
v(t)=e^{{3b_0}(t-t_{0})}-\frac{\alpha\rho_{0}}{6b_{0}^2},\label{40}
\end{eqnarray}
which for $t=t_{0}+\frac{1}{{3b_0}}\ln\left
(\frac{\alpha\rho_{0}}{6b_{0}^2}\right)$ becomes zero. By
reparameterizing the initial value of the cosmological time
according to $e^{-3{b_{0}t_{0}}}=\frac{\alpha\rho_{0}}{6b_{0}^2}$,
the evolution of the anisotropic brane universe starts at $t=0$
from a singular state $v(t=0)=0$. Therefore the expansion scalar,
scale factors, mean anisotropy, and decelerating parameter are
given by
\begin{eqnarray}
\Theta(t)=\frac{3b_{0} e^{{3b_{0}}(t-t_{0})}}{
e^{{3b_{0}}(t-t_{0})}-\frac{\alpha\rho_{0}}{6b_{0}^2}},\label{41}
\end{eqnarray}
\begin{eqnarray}
a_i=a_{0i}
\left[e^{{3b_0}(t-t_{0})}-\frac{\alpha\rho_{0}}{6b_{0}^2}\right]^{1/3}
\left[1-\frac{\alpha\rho_{0}}{6b_{0}^2}e^{-{3b_0}(t-t_{0})}\right]^{\frac{2b_0h_i}{\alpha\rho_0}},
\quad i=1,2,3,\label{42}
\end{eqnarray}
\begin{eqnarray}
A(t)=\frac{h^2}{3b_{0}^2 e^{{6b_{0}}(t-t_{0})}},\label{43}
\end{eqnarray}
\begin{eqnarray}
q(t)=\frac{\alpha\rho_{0}}{2b_{0}^2}
e^{-{3b_{0}}(t-t_{0})}-1.\label{44}
\end{eqnarray}
We consider $\lambda=0$ and show that, within the context of the
present model, the extrinsic curvature can be used to account for
the accelerated expansion of the universe. In figure 1 we have
plotted the behavior of the deceleration parameter for different
values of $\gamma$. The behavior of this parameter shows that when
the geometrical energy density is positive and the pressure is
negative the $AdS_{5}$ bulk is not compatible with the expansion
of the universe. Also, this behavior is much dependent on the
range of the values that $ \gamma_{x}$ can take. The use of the de
Sitter bulk with $\rho_{x}>0$ allows us to use the wealth of
available data from the recent measurements to determine limits on
the values of $\gamma_{x}$ in our geometric model. For having an
accelerating universe we distinguish $\gamma_{x}<\frac{2}{3}$. As
mentioned before, $q(t)>0$ corresponds to the standard
decelerating models whereas $q(t)<0$ indicates an accelerating
expansion at late times. Therefore, the universe undergoes an
accelerated expansion at late times in the absence of a positive
cosmological constant.  As has been noted in \cite{24}, it should
be emphasised here too that the geometrical approach considered
here is based on three basic postulates, namely, the confinement
of the standard gauge interactions, the existence of quantum
gravity in the bulk and finally, the embedding of the brane world.
All other model dependent properties such as warped metric, mirror
symmetries, radion or extra scalar fields, fine tuning parameters
like the tension of the brane and the choice of a junction
condition are left out as much as possible in our calculations.

To understanding the behavior of the mean anisotropy parameter in
our model, let us consider it as a function of the volume scale
factor
\begin{eqnarray}
A(v)=\frac{3h^2}{9\beta^{2}v^{4/3}+3\lambda v^2+3\alpha
\rho_{0}v^{2-\gamma} +9b_{0}^{2}v^{2-\gamma_{x}}+C}.\label{45}
\end{eqnarray}
The behavior of the anisotropy parameter at the initial state
depends on the values of $\gamma$ and $\gamma_{x}$. For an
accelerating universe we  obtain $\gamma_{x}<\frac{2}{3}$. From
equation (\ref{45}), in the limit $v\rightarrow 0$ (singular
state) and taking $\gamma_{x}<\frac{2}{3}$, we find
\begin{eqnarray}
\lim_{v\rightarrow 0} A(v)&=&\frac{3h^2}{C},\hspace{.5
cm}1\leq\gamma\leq 2.\label{46}
\end{eqnarray}
Therefore for a brane world scenario with a confining potential,
the initial state is always anisotropic. In our model the behavior
of the anisotropy parameter coincides with the standard $4D$
cosmology and is different from the brane world models where  a
delta-function in the energy-momentum tensor is used \cite{39} to
confine the mater on the brane. The behavior of the mean
anisotropy parameter of the Bianchi type I and V geometries are
illustrated, for $\gamma_{x}=0.3$ and different values of
$\gamma$, in figure 2. The behavior of this parameter shows that
the universe starts from a singular state with maximum anisotropy
and  ends up in an isotropic de Sitter inflationary phase at late
times. In figure 3 we have plotted the deceleration parameter and
the anisotropy parameter of the Bianchi type I geometry for
$\gamma=1$ and different values of $\gamma_{x}=0,0.3,0.5$.

At this point it would be appropriate to compare our model with
other forms of dark energy such as the $4D$ quintessence. One may
consider a $4D$ effective Lagrangian whose variation with respect
to $g_{\mu\nu}$ would result in the dynamical equations (\ref{4})
compatible with the embedding and with the confined matter
hypotheses \cite{24}. In contrast to the standard model, our model
corresponds to an Einstein-Hilbert Lagrangian which is modified by
extrinsic curvature terms. The resulting Einstein equations are
thus modified by the term $Q_{\mu\nu}$. Since, as was mentioned
before, this quantity is independently conserved, there is no
exchange of energy between this geometrical correction and the
confined matter source. Such an aspect has one important
consequence however; if $Q_{\mu\nu}$ is to be related to dark
energy, as we did in this paper, it does not exchange energy with
ordinary matter, like the coupled quintessence models \cite{40}.
The coupled scalar field models may avoid the cosmic coincidence
problem with the available data being used to fix the
corresponding dynamics and, consequently, the scalar field
potential responsible for the present accelerating phase of the
universe.
\begin{figure}
\centerline{\begin{tabular}{ccc}
\epsfig{figure=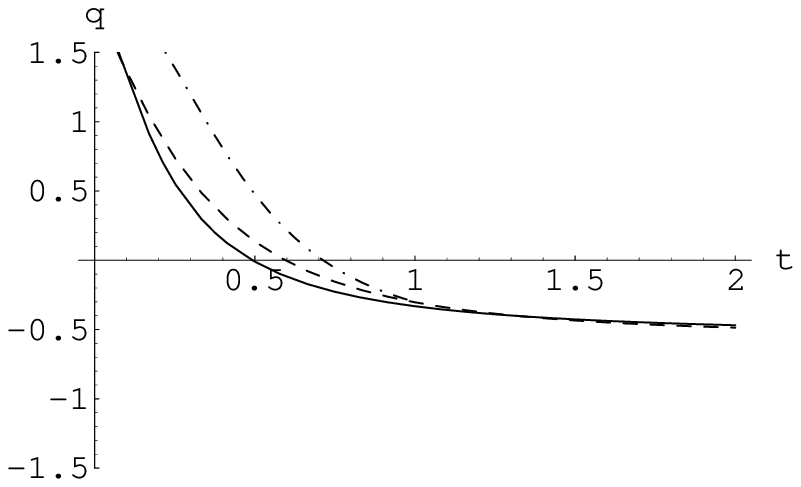,width=6cm}\hspace{20mm}\epsfig{figure=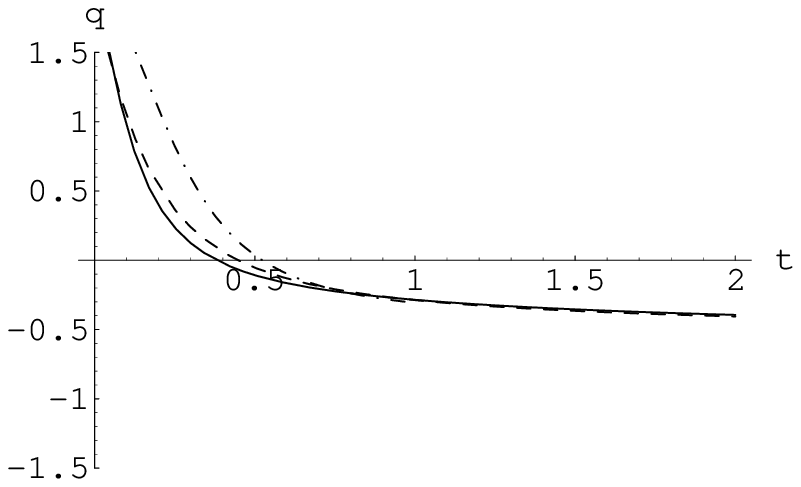,width=6cm}
\end{tabular} } \caption{\footnotesize  Left, the deceleration
parameter of the Bianchi type I universe and right, the same
parameter in the Bianchi type V universe for the de Sitter bulk
with $\gamma=1$ (solid line), $\gamma=4/3$ (dashed line),
$\gamma=2$ (dot-dashed line), $\gamma_{x}=0.3$ and
$\lambda=0$.}\label{fig1}
\end{figure}
\begin{figure}
\centerline{\begin{tabular}{ccc}
\epsfig{figure=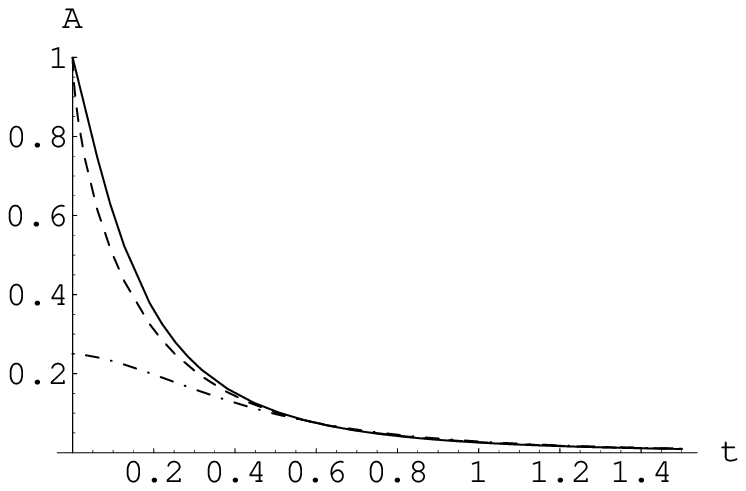,width=6cm}\hspace{20mm}\epsfig{figure=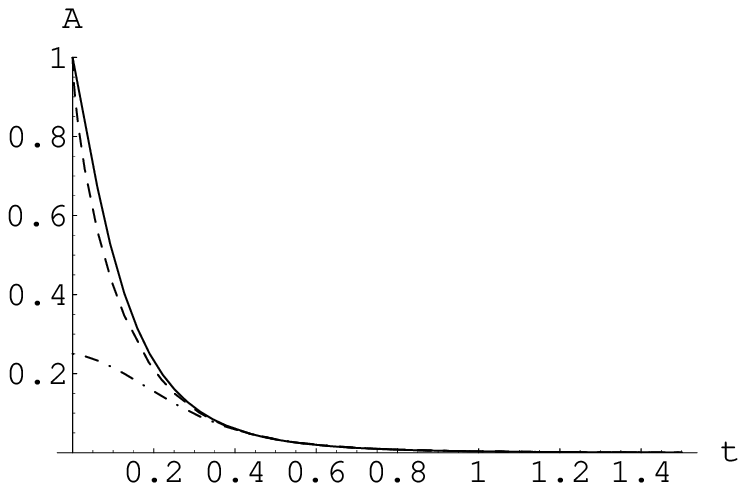,width=6cm}
\end{tabular} }
\caption{\footnotesize Left, the anisotropy parameter of the
Bianchi type I universe and right, the same parameter in the
Bianchi type V universe for the de Sitter bulk with $\gamma=1 $
(solid line), $\gamma=4/3$ (dashed line), $\gamma=2$ (dot-dashed
line), $\gamma_{x}=0.3$ and $\lambda=0$.} \label{fig2}
\end{figure}
\begin{figure}
\centerline{\begin{tabular}{ccc}
\epsfig{figure=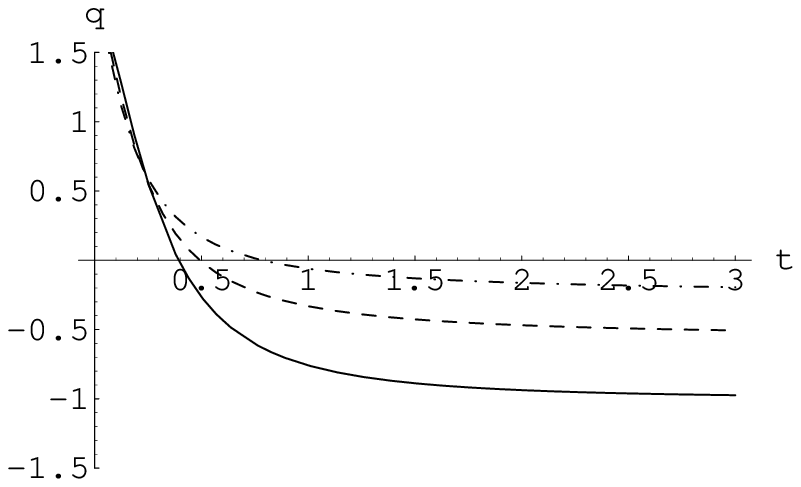,width=6cm}\hspace{20mm}\epsfig{figure=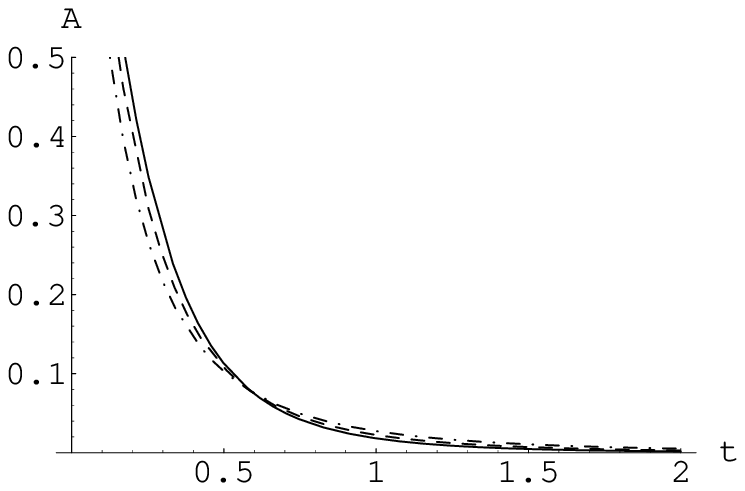,width=6cm}
\end{tabular} }
\caption{\footnotesize Left, the deceleration parameter of the
Bianchi type I geometry and right, the anisotropy parameter in the
Bianchi type I geometry for the de Sitter bulk with $\gamma_{x}=0$
(solid line), $\gamma_{x}=0.3$ (dashed line), $\gamma_{x}=0.5$
(dot-dashed line), $\gamma=1$ and $\lambda=0$.} \label{fig2}
\end{figure}\label{fig3}
\section{Conclusions}
In this paper, we have studied an anisotropic brane world model in
which the matter is confined to the brane through the action of a
confining potential, rendering the use of any junction condition
redundant. We have shown that in an anisotropic brane world
embedded in a constant curvature $dS_{5}$ bulk the accelerating
expansion of the universe can be a consequence of the extrinsic
curvature and thus a purely geometrical effect.  The study of the
behavior of the anisotropy parameter shows that in our model the
universe starts as a singular state with maximum anisotropy and
reaches, for both Bianchi type I and V space-times, an isotropic
state in the late time limit. The study of this scenario in an
inhomogeneous brane will be the subject of a future investigation.

\end{document}